\input harvmac
\input epsf

\let\includefigures=\iftrue
%
%
%
\newfam\black

\input epsf
\noblackbox
%
%
\includefigures
\message{If you do not have epsf.tex (to include figures),}
\message{change the option at the top of the tex file.}
\def\figin{\epsfcheck\figin}\def\figins{\epsfcheck\figins}
\def\epsfcheck{\ifx\epsfbox\UnDeFiNeD
\message{(NO epsf.tex, FIGURES WILL BE IGNORED)}
\gdef\figin##1{\vskip2in}\gdef\figins##1{\hskip.5in}
\else\message{(FIGURES WILL BE INCLUDED)}%
\gdef\figin##1{##1}\gdef\figins##1{##1}\fi}
\def\DefWarn#1{}

\def\figinsert{\goodbreak\midinsert}
\def\ifig#1#2#3{\DefWarn#1\xdef#1{fig.~\the\figno}
\writedef{#1\leftbracket fig.\noexpand~\the\figno}%
\figinsert\figin{\centerline{#3}}\medskip\centerline{\vbox{\baselineskip12pt
\advance\hsize by -1truein\noindent\footnotefont{\bf
Fig.~\the\figno:} #2}}
\bigskip\endinsert\global\advance\figno by1}
\else
\def\ifig#1#2#3{\xdef#1{fig.~\the\figno}
\writedef{#1\leftbracket fig.\noexpand~\the\figno}%
\global\advance\figno by1} \fi

\def\alpalp{\hbox{$\alpha$\kern -0.55em $\alpha$}}
\def\betbet{\hbox{$\beta$\kern -0.50em $\beta$}}

\def\lroverarrow#1{\raise4.2truept\hbox{$\displaystyle
\leftrightarrow\atop\displaystyle#1$}}
\def\underarrow#1{\vbox{\ialign{##\crcr$\hfil\displaystyle
 {#1}\hfil$\crcr\noalign{\kern1pt\nointerlineskip}$\longrightarrow$\crcr}}}

%

\font\teneurm=eurm10 \font\seveneurm=eurm7 \font\fiveeurm=eurm5
\newfam\eurmfam
\textfont\eurmfam=\teneurm \scriptfont\eurmfam=\seveneurm
\scriptscriptfont\eurmfam=\fiveeurm

 \font\teneusm=eusm10 \font\seveneusm=eusm7 \font\fiveeusm=eusm5
\newfam\eusmfam
\textfont\eusmfam=\teneusm \scriptfont\eusmfam=\seveneusm
\scriptscriptfont\eusmfam=\fiveeusm

\font\tencmmib=cmmib10 \skewchar\tencmmib='177
\font\sevencmmib=cmmib7 \skewchar\sevencmmib='177
\font\fivecmmib=cmmib5 \skewchar\fivecmmib='177
\newfam\cmmibfam
\textfont\cmmibfam=\tencmmib \scriptfont\cmmibfam=\sevencmmib
\scriptscriptfont\cmmibfam=\fivecmmib


%

%

%
\def\p{\partial}

\def\Z{{\rm Z}}

\def\R{{\rm R}}

\def\K{{\rm K}}

\def\B{{\rm B}}

\def\CD{{\cal D}}

\def\BZ{\Bbb{Z}}

\def\BFS{{\bf S}}

\def\rmp#1#2#3{Rev. Mod. Phys. {\bf #1} (#2) #3}

\overfullrule=0pt

\Title{} {\vbox{\centerline{Vortex Operator and BKT Transition in Abelian Duality}
}}
\smallskip
\centerline{Tong Chern$^\dagger$}
\smallskip
\centerline{\it School of Science, East China Institute of Technology, Nanchang 330013, China}
\medskip \centerline{\it $^\dagger$ towntong@gmail.com}

\bigskip

\smallskip
\smallskip
\input amssym.tex

\noindent

We give a new simple derivation for the sine-Gordon description of Berezinskii-Kosterlitz-Thouless(BKT) phase transition
(is driven by vortices). Our derivation is simpler than traditional derivations. Besides, our derivation is a continuous field theoretic derivation by using path integration, different from the traditional derivations which are based on lattice theory or based on Coulomb gas model. Our new derivation rely on Abelian duality of two dimensional quantum field theory. By utilizing this duality in path integration, we find that the vortex configurations are naturally mapped to exponential operators in dual description, these operators are the vortex operators that can create vortices, the sine-Gordon description then naturally follows. Our method may be useful for the investigation to the BKT physics of superconductors.

\Date{November, 2014}

\listtoc \writetoc

\newsec{Introduction}
\seclab\intro

\nref\B{V. B. Berezinskii, Sov. Phys. JETP 32, 493 (1971).}\nref\KT{J. M.
Kosterlitz, D. J. Thouless, Journal of Physics C: Solid
State Physics, Vol. 6 pages 1181-1203 (1973).}\nref\K{J. M.
Kosterlitz, J. Phys. C 7, 1046 (1974).}
\nref\MerminWagner{N. D. Mermin, H. Wagner, Phys. Rev. Lett. 17, 1133-1136 (1966).}
\nref\Hohenberg{P. C. Hohenberg, Phys. Rev. 158, 383 (1967).}
\nref\Coleman{Sidney Coleman, Commun. Math. Phys. 31, 259 (1973).}
\nref\rev{For a recent review see "40 years of Berezinskii-Kosterlitz-Thouless theory" edited by Jorge V. Jos¨¨, World Scientific (2013).}
\nref\natured{V. J. Emery and S. A. Kivelson, Nature 374, 434 (1995).}
\nref\rep{V. M. Loktev, R. M. Quick, and S. G. Sharapov, Phys. Rep.,
349, 1 (2001).}
\nref\naturea{Z. Hadzibabic et al., Nature 441, 1118 (2006).}
\nref\natureb{C.-L. Hung, X. Zhang, N. Gemelke, and C. Chin, Nature 470, 236 (2011).}
\nref\naturec{M. Feld, B. Frohlich, E. Vogt, M. Koschorreck, and M. Kohl,
Nature, 480, 75 (2011).}
\nref\rmp{A. Posazhennikova, Rev. Mod. Phys., 78, 1111 (2006).}
\nref\cmft{A. Altland, B. Simons, Condensed Matter Field Theory, Cambridge University Press 2010.}
\nref\WXG{Xiao-Gang Wen, Quantum Field Theory of
Many-body Systems: From the Origin of Sound to an Origin of Light
and Electrons, Oxford University Press 2004.}
\nref\nagaosa{N. Nagaosa, Quantum Field Theory in Condensed Matter Physics, Springer-Verlag Berlin Heidelberg 1999.}
\nref\min{P. Minnhagen, Rev. Mod. Phys. 59, 1001 (1987)}
\nref\benfattoz{L. Benfatto, C. Castellani, and T. Giamarchi, arXiv:1201.2307}
\nref\witten{P. Deligne, D. Kazhdan, e.t.c, Quantum Fields and Strings: A Course For
Mathematicians, Volume II, American Mathematical Society, Institute
for Advanced Study.}
\nref\Benfatto{L. Benfatto, C. Castellani, and T. Giamarchi, Phys. Rev.
Lett. 98, 117008 (2007).}
\nref\benfattob{L. Benfatto, C. Castellani, and T. Giamarchi, Phys. Rev. Lett. 99, 207002 (2007)}
\nref\benfattoc{L. Benfatto, C. Castellani, and T. Giamarchi, Phys. Rev. B 77, 100506(R) (2008)}
\nref\ctb{working in progress}
\nref\prla{V. Schweikhard, S. Tung, and E. A. Cornell, Phys. Rev. Lett.
99, 030401 (2007).}
\nref\prlb{P. Kruger, Z. Hadzibabic, and J. Dalibard, Phys. Rev. Lett. 99,
040402 (2007).}
\nref\prlc{P. Clade et al., Phys. Rev. Lett. 102, 170401 (2009).}
\nref\prld{S. Tung et al., Phys. Rev. Lett. 105, 230408 (2010).}
\nref\prle{T. Yefsah et al., Phys. Rev. Lett. 107, 130401 (2011).}
\nref\prlf{J.-y. Choi, S. W. Seo, W. J. Kwon, and Y.-i. Shin, Phys. Rev.
Lett. 109, 125301 (2012).}
\nref\prlg{J.-y. Choi, S. W. Seo, and Y.-i. Shin, Phys. Rev. Lett. 110,
175302 (2013).}
\nref\prlh{L.-C. Ha et al., Phys. Rev. Lett. 110, 145302 (2013).}
\nref\prli{R. Desbuquois et al., Phys. Rev. Lett. 113, 020404 (2014).}
\nref\praa{S. P. Rath et al., Phys. Rev. A 82, 013609 (2010).}
\nref\prab{T. Plisson et al., Phys. Rev. A 84, 061606 (2011).}
\nref\prac{R. Desbuquois et al., Nat. Phys. 8, 645 (2012).}
\nref\prad{R. J. Fletcher et al., arXiv:1501.02262.}
\nref\giab{M. A. Cazalilla, A. Iucci, and T. Giamarchi, Phys. Rev.
A 75, 051603(R) (2007).}
\nref\poly{A. M. Polyakov, Gauge Fields and Strings, Harwood Academic Publishers 1987.}
\nref\pol{J. Polchinski, String Theory, Cambridge University Press, Cambridge,
UK (1998).}

The Berezinskii-Kosterlitz-Thouless (BKT) \B\KT\K\
phase transition is the unique transition where vortices plays the main role.
This transition can be illustrated by the two dimensional(2D) XY model.
It is well known that this model has no
spontaneously symmetry broken\MerminWagner\Hohenberg\Coleman .
But it does exhibit a phase transition
without the appearance of a spontaneous magnetization. Its low temperature phase (below a critical
temperature $T_c$) contains massless spin waves, and has an
algebraic long range order. Its high temperature phase ($T>T_c$) is
completely disordered by the Coulomb gases of vortices, leads to an exponential decay of spin-spin
correlations, and only short range order can exist.
The BKT transition has been argued to be relevant to many phenomena\rev ,
for examples, to part of the phase diagram of high Tc superconductors\natured\rep ,
to the vortex physics of ultracold-atom \naturea\natureb\naturec , and to weakly interacting bose gases in 2D
(see \rmp for a review).

It is well known that the BKT transition is beyond ordinary Landau paradigm of
continuous phase transitions (which relates phase transition to symmetry breaking).
As a result, people traditionally study the BKT transition,
not by using continuous field theory, but as a lattice theory (the 2D XY spin model) or as an
equivalent Coulomb plasma system\KT\K .
But the BKT transition can also be described by an effective field theory,
the two-dimensional sine-Gordon model \cmft\WXG.
The traditional derivations for this effective field theory description
are via a complicated dual transformation in lattice theory \nagaosa\
or through the Coulomb gas model \cmft\min\benfattoz .
In this paper, we try to give a direct continuous field
theoretic derivation, by further developing the
path-integral formulation of the Abelian-duality of two dimensional quantum field theory \witten .
Our new derivation is elegant and simpler than traditional derivations.
In our derivation, by using path integration, the vortex configurations can be naturally mapped
to exponential operators (the vortex operators) in dual description.
Then, the vortex and anti-vortex gases are naturally mapped to a cosine interaction term,
the sine-Gordon description of BKT physics thus follows.

Since the sine-Gordon description has been used to investigate
the BKT physics of quasi-two-dimensional superconducting systems \benfattoz ,
such as the layered superconductors \Benfatto , the two-dimensional superconductors at finite magnetic
field \benfattob , and to investigate the energy needed to create a vortex core in bilayer films of cuprates \benfattoc ,
we can naturally expect that our new method may cast some lights on these problems\ctb . On the other hand, most ultracold-atom experiments on BKT
physics of vortices \naturea\natureb\naturec\prla-\prad\ are concentrated on interacting Bose gas
in a two dimensional harmonic trap, in contrast to the infinite uniform system,
in these cases our new method may be
much more convenience for theoretical analysis than usual methods \ctb\giab .

The present paper will be organized as following, in next section, we'll give our new simple derivation for the sine-Gordon description of
BKT transition (this is the content of subsection 2.3), to achieve this,
we will firstly develop the Abelian duality and the vortex operator description for vortices,
this is the contents of subsection 2.2 and 2.3. In section three, we'll revisit the BKT transition
based on the sine-Gordon model,
a detail renormalization group analysis will be included in for completion.
In last section, we summarized our results and give an outlook for further developments.

\newsec{Vortex Operator and sine-Gordon Description of BKT Transition}
\seclab\FKT \subsec{Continuous Field Theory Description for BKT Transition} \subseclab\F

In this subsection, we'll develop a continuous field theoretic description for BKT transition by using path integration.
We'll begin from the two-dimensional XY model to set our conventions,
but our description is essentially based on continuous field theory
and does not rely on the lattice theory of XY model.

The two-dimensional XY model is a system of spins constrained to
rotate in the plane of the lattice with spacing $a$. This model is
described by the partition function\poly \eqn\XY{\eqalign{{\cal
Z}&=\int_{-\pi}^{\pi}\prod_{x}{dA_{x}\over 2\pi}\exp[{\beta\over
2\pi}\sum_{x,\delta}(\BFS_x\cdot \BFS_{x+\delta}-1)],\cr
&=\int_{-\pi}^{\pi}\prod_{x}{dA_{x}\over 2\pi}\exp[{\beta\over
2\pi}\sum_{x,\delta}(\cos(A_x-A_{x+\delta})-1)].}} Here $x$ denotes
a site of the 2D lattice, $\delta$ is a unit vector connecting this
site with one of its nearest neighbors, and $\beta=1/T$ is the
inverse temperature. Where $A_x\sim
A_x+2\pi$ is the angle that the $x$-th spin makes with some arbitrary
axis, and we have normalised $|\BFS_x|=1$.

Since we only interest the long-distance behaviors of this system, we can
develop a continuous field theoretic description for it, in this
description, only slowly varying configurations will give
significant contributions to the partition function so that we may
expand the action up to terms quadratic in the angles $A_x$
\eqn\action{S={1\over 4\pi
T}\sum_{x,\delta}(A_x-A_{x+\delta})^2\approx {1\over 4\pi
T}\int_{\R^2}\parallel dA\parallel^2} where $A(x)\sim A(x)+2\pi$ is a continuous scalar field,
$dA={\p A\over\p x^i}dx^i$ is the wedge
differential of $A$, and $\parallel d
A\parallel^2=(dA)\wedge\ast(dA)$, here $\ast$ is the Hodge star operator
of two dimensions, and $\wedge$ stands for the wedge product of differential forms.
If we ignore the periodicity of the angular
variable $A$ at first(hence we ignore vortex configurations), then
the partition function of the system can be given as an appropriate
path integral of $A$ field, with weight $W[A]$,
\eqn\WA{W[A]=\exp\left[-{1\over 4\pi T}\int_{\R^2}\parallel
dA\parallel^2\right].}

To account for the periodicity of $A$, we must sum over certain
singular configurations (in
the continuous field theoretic description) carrying nontrivial topological numbers. That is, we
should allow field $A$ to be a multi-valued function, so that it may have
$2\pi$ jumps at certain branch cuts. These field configurations stand for the
vortex excitations.

Now we'll describe how to handle the vortex configurations,
in the $A$ field path integration. When there is a vortex singularity at point $p$, we draw an infinite
small circle $C_p$ around $p$ counterclockwise, cut out the inner of
$C_p$ -- to get rid of the vortex singularity -- and replace the
path integration over the field configuration on this inner patch by
a small constant $g$, which stands for the contribution of the inner of
a vortex. After that we take the configuration of $A$ along
$C_p$ (this configuration is determined by the vortex located at
$p$) as the boundary condition for the $A$ field configuration at
the outside of $C_p$, and we'll perform the path integral at the
outside of $C_p$ with this vortex boundary condition. Thus the total
contribution of a vortex configuration, to the path integral, is the
multiplication of this outer patch path integration and the inner patch contribution $g$.

In fact, for a $n$-vortex, located at $p$, carrying topological
quantum number(winding number) $n\in \BZ$, we have \eqn\tqn{\int_{C_p}{F_A\over
2\pi}=n,} where $F_A=dA$. Thus, for this vortex, the $A$ field boundary condition along $C_p$
can be simply described as: the value of $A$
will increase $n2\pi$ after walking around $C_p$ a
circle counterclockwise. If $n$ is negative, we will call this $n$-vortex
as an anti-vortex.

For the vortex with topological winding number 2, 2-vortex, we can
image that it is composed of two tightly bound 1-vortices. However, because vortex-vortex interaction is
always mutually repulsion as we will see in subsection 2.3, the two composite 1-vortex always tend
to separate from each other, thus the unbound configuration have lower
energy and higher weight in the path integration. The same
arguments can also be applied to vortices and anti-vortices with higher
winding numbers. Therefore, we can ignore the contributions, to the
partition function, of all higher vortices with topological quantum
numbers greater than one (or anti-vortices with winding numbers
smaller than minus one), but only include the contributions of well
separated 1-vortices, and 1-anti-vortices ($-1$-vortices), they form
a kind of dilute gas.

For each dilute gas with $m$ 1-vortices and $n$ $-1$-vortices, we
can perform the corresponding path integral, which is a
generalization of the path integration with single vortex, that we
have described the prescription for it. Finally, we will sum over the
contributions of all the dilute gases with different numbers of vortices and anti-vortices,
including the contributions with no vortices, end up
with a 1-vortices approximate partition function $Z_{1vortexes}$,
\eqn\summn{Z_{1vortexes}=\sum_{m,n}{1\over m!}{1\over
n!}g^{m+n}\int_{m. n}[\CD A]\exp\left[-{1\over 4\pi
T}\int_{\R^2}\parallel dA\parallel^2\right],} where $\int_{m,n}[\CD
A]$ stands for the path integration over the $A$ field
configurations at the outside of each 1-vortex and
of each $-1$-vortex, with
corresponding vortex boundary conditions at the infinite small circles. Noticing that the position coordinates
of each vortex and each anti-vortex must be integrated too, since the vortices located at different
positions stand for different configurations. The factors ${1\over
m!}$ and ${1\over n!}$ in \summn\ come from the permutations of identical
vortices and identical anti-vortices respectively, and the factor $g^{m+n}$ comes from the inner configurations
of these $m$ vortices and $n$ anti-vortices. Where we have use the symbol
$Z_{1vortexes}$, to remind us that we have made 1-vortices (and
$-1$-vortices) and dilute gas approximation.

In the following two subsections, we'll carry out the path integral
in \summn, by developing the Abelian duality \witten\
and vortex operator description to the vortices.

\subsec{Abelian Duality} \subseclab\duality

In this subsection, instead of the $A$ field description to BKT transition
that we have described in \F,
we'll develop a dual description in terms of
some $B$ field by utilizing the two dimensional Abelian duality\witten . The main idea is, in
short, to transform the vortices in $A$
field description to some vortex operators in dual description.

We begin from the configurations without vortices, by noticing that the
path integration $\int[\CD A] W[A]$ is equivalent to
\eqn\ATB{Z=\int[\CD A \CD B]\exp\left[{i\over 2\pi}
\int_{\R^2}(dB\wedge F_A)\right]W[A], } where $F_A=dA$, and we have
normalized field $B$ to have periodicity $B\sim B+2\pi$. This
equivalence can be seen by noticing that the exponential
factor in \ATB\ is automatically trivial by integrating by part,
thus the path integral over $B$ only contribute a trivial infinite
constant, which can be regularized easily.

Now we exchange the order of $A$, $B$ functional integrations. Note
that the functional integration over $A$ is basically the same as the
functional integration over $F_A=dA$, since the $A$ field configurations are topologically trivial.
Therefore, we firstly perform the
integration over $dA$ in \ATB,
the result, ignoring a constant factor that comes from the gaussian integration over $dA$, is \eqn\ZB{Z=\int[\CD
B]\exp\left[-{T\over 4\pi}\int\parallel dB\parallel^2\right].}
In other words, the partition function $\int[\CD
A]W[A]$(without vortices) is completely equivalent to \ZB, a functional
integration over a dual field $B$.

To become familiar with this duality transformation, we'd like to work out another example,
we'll calculate the path integral $\int[\CD
A](dA)W[A]$, with insertion $dA$.
By using the duality transformation as above, we can see that $\int[\CD
A](dA)W[A]$ is equivalent to
\eqn\BInsert{\int[\CD B](iT\ast dB)\exp\left[-{T\over 4\pi}\int
\parallel dB\parallel^2\right].}

With these preparations, we can now reexpress the $A$ field path
integration with 1-vortex boundary conditions as the dual $B$ field
path integration with appropriate local operator insertions,
these local operators are the vortex operators that can create vortices.
These are the content of next subsection.

\subsec{Vortex Operator and sine-Gordon Description}

\noindent

We can now derive a dual $B$ field description for $Z_{1vortexes}$ \summn ,
it turns out to be the sine-Gordon description for BKT transition.
To achieve this, we firstly calculate the contribution of a single
1-vortex.

We recall that, in $A$ field
description, a single 1-vortex located at $p$ is a vortex boundary
with corresponding boundary condition at the infinite small circle
$C_p$ (the inner of $C_p$ has been cut out and replaced by a
factor $g$). After walking along $C_p$ a circle counterclockwise,
the value of $A$ will increase $2\pi$. We'll denote the $A$
field path integral with this vortex boundary condition as $Z[C_p]$.

To proceed, we'll denote the space at the outside of $C_p$ as
$C\times\R^+$, and will pick a right hand orientation on it. Obviously,
as the boundary of $C\times\R^+$, the considered 1-vortex boundary
has a correlated clockwise orientation, we'll denote this boundary
as $C^-_p$, the superscript $-$ is used to indicate its clockwise orientation.
We then consider the exponential factor
\eqn\ef{\exp\left[{i\over 2\pi}\int_{C\times\R^+}dB\wedge
F_A\right].} One can rewrite this factor as
$\exp\left[{i\over 2\pi}\int_{C\times\R^+}d(B F_A)\right]$, which
can be integrated (at the boundary $C^-_p$) as
\eqn\effB{\exp\left[iB(p)\int_{C^-_p}{F_A\over
2\pi}\right],} since $C^-_p$ is an infinite
small circle. By noticing that $C^-_p$ is the vortex boundary
of the considered 1-vortex, but with clockwise orientation,
we have $\int_{C^-_p}{F_A/
2\pi}=-\int_{C_p}{F_A/
2\pi}=-1$, thus \eqn\efB{\exp\left[iB(p)\int_{C^-_p}{F_A\over
2\pi}\right]=\exp\left[-iB(p)\right].}
Obviously, one can cancel this final result by multiplying an
additional factor $\exp\left[iB(p)\right]$ to the exponential \ef,
thus \eqn\e{\exp\left[iB(p)\right]\exp\left[{i\over
2\pi}\int_{C\times\R^+}dB\wedge F_A\right]=1.}

From these discusses, we can see that the $A$ field path integral
$Z[C_p]$ is equivalent to the following path integral
\eqn\expAB{Z[C_p]=\int[\CD A \CD
B]\exp\left[iB(p)\right]\exp\left[{i\over 2\pi}
\int_{C\times\R^+}(dB\wedge F_A)\right]W[A],} since the path
integration over $B$ is trivially an infinite constant (due to \e)
that can be easily regularized. After exchanging the integral order of $A, B$
functional integrations and carrying out the $A$ field integration, just like we have done in subsection \duality, we can get
(ignoring an irrelevant constant that comes from the gaussian integration
of $dA$) \eqn\ZBC{Z[C_p]=\int[\CD B]\exp[iB(p)]\exp\left[-{T\over
4\pi}\int_{\R^2}\parallel dB\parallel^2\right].}

The above derivation tells us that, in $B$ field description, a
single 1-vortex located at $p$ can be created out by inserting a local
operator $\exp[iB(p)]$ in $B$ field path integral. Likewise, to create a single $-1$-vortex at
$p$, one should insert a local operator $\exp[-iB(p)]$. In general, a $m$-vortex located at $p$ is
created by exponential operator $\exp[imB(p)]$. These exponential operators are the vortex operators.

As an exercise, we can calculate the potential energy $U_{eff}(x)$
and the force $F(x)$ between a $m$-vortex located at $x$ and a
$n$-vortex located at the origin $0$. Obviously, $U_{eff}(x)$ is
determined, in $B$ field description that we just developed, by a
two points correlation function of vortex operators
\eqn\potential{\exp\left[-\beta(U_{eff}(x)+U_L)\right]=\langle
\exp[imB(x)]\exp[inB(0)]\rangle,} where the adjustable constant
$U_L$ is introduced to set $U_{eff}(L)=0$, $L$ denotes the space length of
the considered system. One can easily find $U_{eff}(x)={mn\over
2}\ln(L^2/x^2)$, which is the 2D Coulomb potential between an
electric charge $m$ and and an electric charge $n$. The force $F(x)$
is $F(x)=mn/|x|$, which is repulsion when $m, n$ have the same sign
(two vortices or two anti-vortices), otherwise is attractive (a
vortex and an anti-vortex). This will lead to the Coulomb gas
description of the planar XY model\KT.

We can now reformulate $Z_{1vortexes}$ \summn , by utilizing the
vortex operator descriptions for 1-vortex and $-1$-vortex, as
\eqn\Zmn{\eqalign{Z_{1vortexes}=&\sum_{m,n}{1\over m!}{1\over
n!}g^{m+n}\int[\CD B]\int
d^2x_1d^2x_2...d^2x_m\exp[i\sum^{m}_{i=1}B(x_i)]\cr &\cdot\int
d^2x_1'...d^2x'_n\exp[-i\sum^{n}_{j=1}B(x'_j)]\exp\left[-{T\over
4\pi}\int_{\R^2}\parallel dB\parallel^2\right],}} where $x_i, x'_j$
stands for the positions of $i$-th vortex and $j$-th
anti-vortex respectively. And we have integrated over the position coordinates
of vortices to account for the total contributions of all possible
configurations.

It seems that there are contributions from any configurations with
arbitrary $m$ 1-vortices and arbitrary $n$ $-1$-vortices,
but in fact only the neutral configurations with equal number of 1-vortices and $-1$-vortices
can have nonzero contributions. This can be seen by firstly integrating
out the zero mode of $B$ field in the path integration of \Zmn .
Thus, one need only consider the neutral Coulomb gas of vortices and anti-vortices.

But if we firstly carry out the summation over
non-negative integers $m, n$, in \Zmn, we can reexpress $Z_{1vortexes}$
as \eqn\ZI{Z_{1vortexes}=\int[\CD B]\exp\left[-{T\over
4\pi}\int_{\R^2}\parallel dB\parallel^2+g\int_{\R^2}\cos(B)\right].}
This is a two-dimensional sine-Gordon model, the well known effective field
theory description \cmft\WXG\ for BKT phase transition.
Thus, we have arrived at our mainly purposes of
the present paper, that is to provide an elegant new derivation for the
sine-Gordon description for BKT transition,
our derivation is directly based on continuous field theory rather than
based on the lattice theory or based on Coulomb gas model.

\newsec{BKT Phase Transition Revisit}

\noindent

We now revisit the BKT phase transition, by utilizing the sine-Gordon description \ZI.
All the results of this section can be compared to the literatures\KT\K\cmft\WXG . From
the expression \ZI\ of $Z_{1vortexes}$, one can see that, at
extremely low temperature, $B$ field fluctuations are very large, these
fluctuations will erase the contributions of vortices. Thus, at extremely low temperature we can
completely ignore vortices and anti-vortices. The
spin-spin correlation can then be easily calculated within $A$ field
description \WA
\eqn\twopoint{\langle\exp[iA(x)]\exp[-iA(0)]\rangle\sim\left({1\over
x^2}\right)^{T/2}.} While at high temperature, phase rigidity of \ZI is
large, the cosine term becomes important, thus field $B$ is classically fixed at $0$.
In this case one can expanse field $B$
around $0$ in \ZI , to see that the system is gaped, and the order of the
system is short range.

To determined the critical temperature $T_{c,g=0}$ at $g=0$, we'll view the
effective theory \ZI\ as a $g\cos(B)$ perturbation to the free field
theory with action $S_0={1\over 4\pi\beta}\int_{\R^2}||dB||^2$, the
critical temperature is at the value to make the $\cos(B)$
deformation marginal, hence the conformal weight
$\Delta=\beta/2$ of $\cos(B)$ becomes $2$. This gives us
\eqn\criticalt{1/(2T_{c,g=0})=2,\  T_{c,g=0}=1/4.}

Furthermore, at the vicinity of $g=0$, one can calculate the renormalization group(RG) flow
of the dimensionless coupling constant $y=ga^2$ (to first order of
$y$), this flow should be proportional to $x=\Delta-2=\beta/2-2$ since
$\Delta=2$ makes $y\cos(B)$ marginal. A detail perturbation
calculation tells us \eqn\yRG{dy=-xyd\ln a.} We can see that, when $T<1/4$, $x>0$,
$y\cos(B)$ is irrelevant, at long distance the RG will flow to a conformal field theory (CFT)
of a free boson, and there is an algebraic long-range correlation
between two $SO(2)$ spins \twopoint. In fact, in this case, the
vortex and anti-vortex excitations will be paired up and combined
into neutral dipoles (due to the Coulomb attraction between them), and the
system is dielectric. Such dipoles only have negligible influence to the
thermodynamical behavior of the system. While, when $T>1/4$, $x<0$, the
vortex perturbation $y\cos(B)$ is relevant, the system has an energy
gap, the RG will flow to trivial at long distance. In this case, the
dipoles will dissociate, form a plasma of vortices and anti-vortices, and a screening
length will be generated dynamically. When $T=1/4$, $x=0$, $y\cos(B)$ is
marginal, if $y=0$, the system will flow to a $\Z_2$ orbifold CFT\pol.

At the vicinity of fixed point $x=\beta/2-2=0$, $y=0$ (hence
$T=1/4$, $g=0$), one can also calculate the RG flow of $x$ to second
order of $y$ and $x$, with the result \eqn\xRG{dx=-Cy^2d\ln a,}
where $C$ is a positive constant which can be set to any positive
number by changing the scale of the length scale $a$. By utilizing
the set of differential equations \yRG\ and \xRG, one can
get a better understanding to the low temperature ($T<T_c$) behaviors
and high temperature ($T>T_c$) behaviors of the system\cmft\WXG. At the critical
temperature $T_c$, the system will flow to the fixed point $x=y=0$,
where the theory is described by a $\Z_2$ orbifold CFT. In $A$ field
description, this CFT can be described as path integral over $W[A]$\WA\
with $T=1/4$. Thus the critical index $\eta$ of the spin-spin correlation
at $T_c$ will be $\eta=1/4$.

As we have stressed, in deriving the effective theory \ZI, we have
made 1-vortex approximation. It is natural to expect
that di-vortices will contribute a similar term $g_2\cos(2B)$
with a much smaller factor $g_2\ll g$. When the temperature is
raised to 1, $\cos(2B)$ will become relevant too. Therefore, our 1-vortex
approximation works only when $T$ is less than $1$.

\newsec{Summary and Outlook}

We have given a simple new derivation for the sine-Gordon description of the BKT phase transition,
by using the path integral formalism of the Abelian duality of two dimension quantum field theories.
Our derivation is directly based on continuous field theory rather than on lattice theory or Coulomb gas model.
In our formulation, the $n$-vortex configuration in $A$ description
is mapped to the vortex operator $\exp[inB(p)]$ in dual $B$ description (where $n\in\BZ$ is the winding number of the $n$-vortex),
and the contributions of neutral vortex anti-vortex gases are naturally mapped to a cosine term $g\int\cos(B)$.
Hence, the sine-Gordon description follows. Based on this effective field theory description,
we revisited the renormalization group analysis of the BKT transition in detail, and we also derived the critical
index $\eta=1/4$ of the spin-spin correlation, our results are consistent with the literatures.

Furthermore, since the sine-Gordon description has been applied to investigate
the BKT physics of quasi-two-dimensional superconductors \benfattoz ,
it is natural to expect that our new method based on path integral formulation of the Abelian duality
and vortex operators
may cast some lights on these problems\ctb .
On the other hand, the likewise Abelian duality may be developed
for two dimensional field theories in a trap\ctb, thus our methods may be used to the theoretical
analysis for recent ultracold-atom experiments on vortex physics of BKT transition\naturea-\naturec\prla-\prli,
since these experiments are concentrated on interacting Bose gas
in a two dimensional harmonic trap, in contrast to the infinite uniform system.

\newsec{Acknowledgements}

The authors acknowledge the support of the Doctoral Startup Package Fund of East
China Institute of Technology (No. DHBK201203).

\listrefs
\end